# ON NONLINEAR DYNAMICS OF THE PENDULUM
# WITH PERIODICALLY VARYING LENGTH

*Anton BELYAKOV, Alexander SEYRANIAN*

a_belyakov@inbox.ru,  seyran@imec.msu.ru

*Institute of Mechanics, Moscow State Lomonosov University,*
*Michurinsky pr. 1, Moscow 119192, RUSSIA*

*Dynamic behavior of a weightless rod with a point mass sliding along the rod axis according to periodic law is studied. This is the pendulum with periodically varying length which is also treated as a simple model of child's swing. Asymptotic expressions for boundaries of instability domains near resonance frequencies are derived. Domains for oscillation, rotation, and oscillation-rotation motions in parameter space are found analytically and compared with numerical study. Two types of transitions to chaos of the pendulum depending on problem parameters are investigated numerically.*

## 1. Introduction

Oscillations of a pendulum with periodically variable length (PPVL) is the classical problem of mechanics. Usually, the PPVL is associated with a child's swing. Everyone can remember that to swing a swing one must crouch when passing through the middle vertical position and straighten up at the extreme positions, i.e. perform oscillations with a frequency which is approximately twice the natural frequency of the swing. Despite popularity of the swing, in the literature, where this problem is referred to in [Kauderer 1958, Bogolyubov, Mitropolsky 1961, Panovko, Gubanova 1987, Magnus 1976, Bolotin 1999], there are not many analytical and numerical results on behavior of the PPVL. In [Kauderer 1958] an approximate relation between the amplitude of a periodic orbit of the pendulum and the excitation frequency was found. In [Bogolyubov, Mitropolsky 1961, Panovko, Gubanova 1987] the PPVL was cited as an example of a periodic system described by an equation different from the Mathieu equation but without further analysis. Change of energy and first instability region for the pendulum with variable piecewise constant length were discussed in [Magnus 1976]. An attempt to find the first instability region for the PPVL was undertaken in [Bolotin 1999] but it is not correct. The stability analysis of the lower vertical position of the pendulum with damping and arbitrary periodic excitation function was carried out in [Seyranian 2004]. In that paper the instability (parametric resonance) regions were found in the form of semi-cones in three-dimensional parameter space.

In papers [Pinsky, Zevin 1999, Zevin, Filonenko 2007] qualitative analysis of periodic solutions and their stability for a PPVL, no damping, and arbitrary excitation amplitude was conducted. In these two papers the adequacy of the PPVL as a model of a swing is also discussed. In paper [Bozduganova, Vitliemov 2009] the variant of PPVL with Coulomb dry-friction at the pivot is studied numerically.

The PPVL is much less studied than the pendulum with oscillating support which is often referred to simply as a *parametrically driven pendulum* or *parametric pendulum*; see e.g. papers [Szemplinska-Stupnicka et al. 2000, Xu et al. 2005, Lenci et al. 2008] and the references therein. These two pendula are described by different analytical models and consequently, possess different dynamical properties. For example, the PPVL cannot be stabilized in the inverted vertical position. Nevertheless, the methods used for dynamical analysis of one pendulum are applicable for the other one. The methodological peculiarity of this work is in the assumption of quasi-linearity of the system which allows us to derive higher order approximations by the averaging method.

The present paper is devoted to the study of regular and chaotic motions of the PPVL. Motivation of this work is to extend authors' earlier results [Seyranian 2004, Seyranian, Belyakov 2008, Belyakov et al. 2008, Belyakov et al. 2009] in investigating dynamics of this rather simple but interesting system.

## 2. Main relations

Equation for motion of the PPVL can be derived with the use of angular momentum alteration theorem; see [Kauderer 1958, Bogolyubov, Mitropolsky 1961, Panovko, Gubanova 1987, Magnus 1976]. Taking into account also linear damping forces, we obtain

(1) $\quad \dfrac{d}{dt}\left(ml^2 \dfrac{d\theta}{dt}\right) + \gamma l^2 \dfrac{d\theta}{dt} + mgl \sin\theta = 0$

where $m$ is the mass, $l$ is the length, $\theta$ is the angle of the pendulum deviation from the vertical position, $g$ is the acceleration due to gravity, and $t$ is the time (Fig. 1).

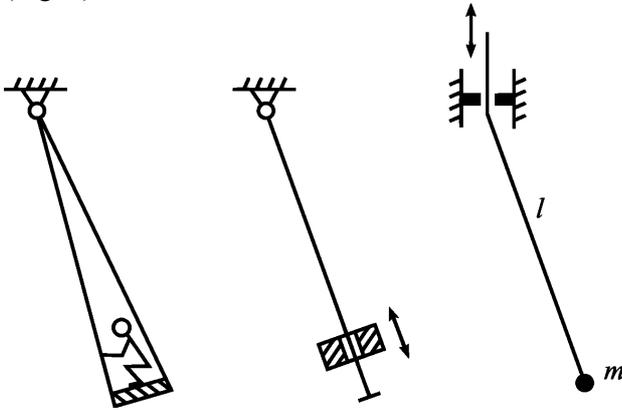

**Fig. 1. Schemes of the pendulum with periodically varying length (PPVL).**

It is assumed that the length of the pendulum changes according to the periodic law

(2) $\quad l = l_0 + a\varphi(\Omega t)$

where $l_0$ is the mean pendulum length, $a$ and $\Omega$ are the amplitude and frequency of the excitation, $\varphi(\tau)$ is the smooth $2\pi$-periodic function with zero mean value.

We introduce the following dimensionless parameters and variables

(3) $\tau = \Omega t,\ \varepsilon = \dfrac{a}{l_0},\ \Omega_0 = \sqrt{\dfrac{g}{l_0}},\ \omega = \dfrac{\Omega_0}{\Omega},\ \beta = \dfrac{\gamma}{m\Omega_0}$

Then, Eq. (1) can be written in the following form

(4) $\quad \ddot{\theta} + \left(\dfrac{2\varepsilon\dot{\varphi}(\tau)}{1+\varepsilon\varphi(\tau)} + \beta\omega\right)\dot{\theta} + \dfrac{\omega^2 \sin\theta}{1+\varepsilon\varphi(\tau)} = 0.$

Here the dot denotes differentiation with respect to new time $\tau$. Behavior of the system governed by Eq. (4) will be studied in the following sections via analytical and numerical techniques depending on three dimensionless problem parameters: the excitation amplitude $\varepsilon$, the damping coefficient $\beta$, and the frequency $\omega$ under the assumption that parameters $\varepsilon$ and $\beta$ are small.

It is convenient to change the variable by the substitution

(5) $\quad \theta = q/(1+\varepsilon\varphi(\tau))$

Using this substitution in Eq. (4) and multiplying it by $1+\varepsilon\varphi(\tau)$ we obtain the equation for $q$ as

(6) $\quad \ddot{q} + \beta\omega\dot{q} - \dfrac{\varepsilon(\ddot{\varphi}(\tau) + \beta\omega\dot{\varphi}(\tau))}{1+\varepsilon\varphi(\tau)}q$
$\quad\quad + \omega^2 \sin\left(\dfrac{q}{1+\varepsilon\varphi(\tau)}\right) = 0.$

This equation is useful for stability study of the vertical position of the pendulum as well as analysis of small oscillations.

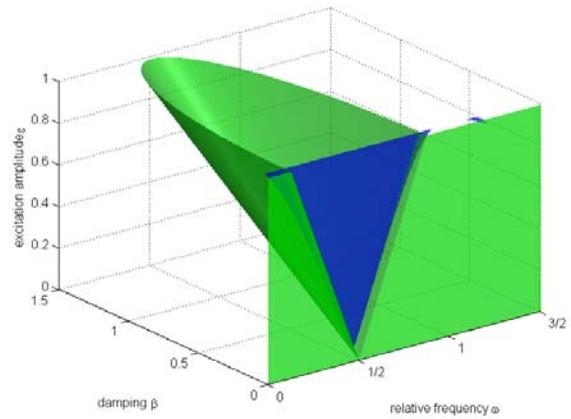

**Fig. 2. Instability region boundary of the PPVL (green surface) in parameter space $(\omega, \beta, \varepsilon)$ in comparison with numerical results in section at $\beta = 0.05$ (blue surface).**

## 3. Instability of the vertical position

Let us analyze the stability of the trivial solution $q = 0$ of the nonlinear equation (6). Its stability with respect to the variable $q$ is equivalent to that of Eq. (4) with respect to $\theta$ due to relation (5). According to Lyapunov's theorem on stability based on a linear approximation for a system with periodic coefficients the stability (instability) of the solution $q = 0$ of Eq. (6) is determined by the stability (instability) of the linearized equation

(7) $\quad \ddot{q} + \beta\omega\dot{q} + \dfrac{\omega^2 - \varepsilon(\ddot{\varphi}(\tau) + \beta\omega\dot{\varphi}(\tau))}{1+\varepsilon\varphi(\tau)}q = 0.$

Expanding the ratio in (7) into Taylor's series about $\varepsilon = 0$ and keeping only first order terms with respect to small parameters $\varepsilon$ and $\beta$ we obtain the following equation

(8) $\quad \ddot{q} + \beta\omega\dot{q} + \left[\omega^2 - \varepsilon(\ddot{\varphi}(\tau) + \omega^2\varphi(\tau))\right]q = 0.$

This is Hill's equation with damping with the periodic function $-\ddot{\varphi}(\tau)-\omega^2\varphi(\tau)$. It is known that instability (i.e. parametric resonance) occurs near the frequencies $\omega = k/2$, where $k = 1, 2, \ldots$ Instability domains in the vicinity of these frequencies were obtained in [Seyranian 2001, Seyranian, Mailybaev 2003] analytically. In three-dimensional space of parameters $\varepsilon$, $\beta$ and $\omega$ these domains are described by half-cones, Fig. 2

(9) $(\beta/2)^2+(2\omega/k-1)^2<(a_k^2+b_k^2)(3\varepsilon/4)^2$,
$\beta \geq 0$,

where $k = 1, 2, \ldots$, $a_k$ and $b_k$ are the Fourier coefficients of the periodic function $\varphi(\tau)$
(10)
$$a_k = \frac{1}{\pi}\int_0^{2\pi}\varphi(\tau)\cos k\tau d\tau, \quad b_k = \frac{1}{\pi}\int_0^{2\pi}\varphi(\tau)\sin k\tau d\tau.$$

Inequalities (9) give us the first approximation of the instability domains of the lower vertical position of the swing. These inequalities were obtained in [Seyranian 2004] using different variables.

For a general system of linear differential equations with periodic coefficients only first order terms in the series of a monodromy matrix with respect to small parameters contribute to the first non-degenerate approximation of instability domains; see [Seyranian, Mailybaev 2003]. That is why to obtain the first order approximation for the instability domains we can omit higher order terms in Eq. (7) and use Eq. (8).

Note that each $k$th resonance domain in relations (9) depends only on the $k$th Fourier coefficients of the periodic excitation function. Particularly, for $\varphi(\tau) = \cos(\tau)$, $k = 1$ we obtain $a_1 = 1$ and $b_1 = 0$. Thus, the first instability domain takes the form
(11) $\beta^2/4+(2\omega-1)^2<9\varepsilon^2/16$, $\beta \geq 0$.

This boundary of the first instability domain ($k = 1$) is presented in Fig. 2 by the green surface demonstrating a good agreement with the numerically obtained instability domain at the section $\beta = 0.05$ (blue surface). The boundary (11) is also drawn in Figs. 5 and 6 by solid white lines. It is easy to see that in (8) for the second resonance domain ($k = 2$, $\omega = 1$) the excitation function is zero for $\varphi(\tau) = \cos(\tau)$. This explains why the second resonance domain is empty, and the numerical results confirm this conclusion; see Fig. 2. Inside the instability domains (9) the vertical position $q = 0$ becomes unstable and motion of the system can be either regular (limit cycle, regular rotation) or chaotic.

## 4. Limit cycle

When the excitation amplitude $\varepsilon$ is small, we can expect that the oscillation amplitude $q$ in Eq. (6) is also small. We suppose that $\varepsilon$ and $\beta$ are small parameters of the same order as well as the factor of nonlinearity. Then, we can expand the sine into Taylor's series around zero in Eq. (6) and keep only

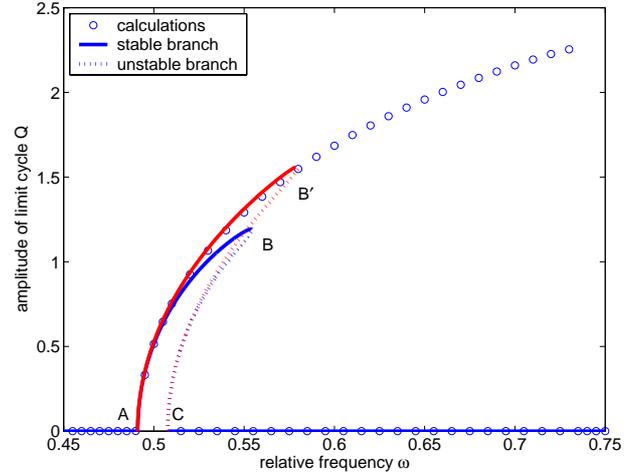

**Fig. 3. Frequency-response curve for the parameters $\varepsilon = 0.04$ and $\beta = 0.05$. Amplitude $Q$ of the limit cycle depending on the relative excitation frequency $\omega$.**

first three terms. Eq. (6) in the first approximation with respect to small parameters takes the following form
(12) $\ddot{q}+\omega^2 q = f(q,\dot{q},\tau)$,

where $f(\cdot)$ is the small function
$$f(q,\dot{q},\tau) = \varepsilon(\ddot{\varphi}(\tau)+\omega^2\varphi(\tau))q+\omega^2\left(\frac{q^3}{6}-\frac{q^5}{120}\right)-\beta\omega\dot{q}$$

because we have assumed that absolute value of the nonlinear term $\omega^2(q^3/6 - q^5/120)$ in Eq. (12) is much less than the absolute value of the linear term $\omega^2 q$. We study the parametric excitation of nonlinear system (12) with the periodic function $\varphi(\tau) = \cos(\tau)$ at the first resonance frequency $\omega = 1/2$. We are looking for an approximate solution of system (12) in the form $q = a(\tau)\cos(\tau/2+\psi(\tau))$ using the averaging method for the resonance case, where $a(\tau)$ and $\psi(\tau)$ are the slow amplitude and phase. As a result, we get the system of first order differential equations for the slow variables $a$ and $\psi$

(13)
$$\dot{a} = -\frac{\sin(\tau/2+\psi)}{\omega}\times$$
$$\times f(a\cos(\tau/2+\psi),-a\omega\sin(\tau/2+\psi),\tau)$$
$$\dot{\psi} = \omega-\frac{1}{2}-\frac{\cos(\tau/2+\psi)}{a\omega}\times$$
$$\times f(a\cos(\tau/2+\psi),-a\omega\sin(\tau/2+\psi),\tau)$$

System (13) has small right hand sides because we assume that $\omega \approx 1/2$ and function $f(\cdot)$ is small. To obtain the first order approximation we average right hand sides of system (13) with respect to time $\tau$ over the period $4\pi$. As a result we obtain the system of averaged differential equations for corresponding averaged variables $Q$ and $\Psi$ instead of $a$ and $\psi$. If we set time derivatives to zero in this system, then it yields two conditions for steady solutions. Excluding phase $\Psi$ from these conditions and disregarding the trivial solution $Q = 0$ we derive the following transcendental equation

(14)
$$\frac{\beta^2\omega^2}{\left(1-\omega^2\left(1-\frac{Q^2}{12}+\frac{Q^4}{384}\right)\right)^2} + \left(\frac{\frac{1}{2}-2\omega^2\left(1-\frac{Q^2}{8}+\frac{Q^4}{192}\right)}{1-\omega^2\left(1-\frac{Q^2}{6}+\frac{Q^4}{128}\right)}\right)^2 = \varepsilon^2$$

This equation gives frequency-response curve AB'C (red line in Fig. 3) which better coincides with numerical results (circles) than curve ABC obtained in [Seyranian, Belyakov 2008]. As can be seen in Fig. 3, the stable branch AB' of curve AB'C is in good agreement with the results of numerical modeling up to the amplitudes $Q \approx \pi/2$.

### 5. Regular rotations

We will say that the system performs regular rotations if a nonzero average rotational velocity exists:

$$b = \lim_{T \to \infty} \frac{1}{T} \int_0^T \dot{\theta} d\tau.$$

Velocity $b$ is a rational number because regular motions can be observed only in resonance with the excitation. Motion with fractional average velocity such as $b = 1/2$ in Fig. 4(a) is usually called oscillation-rotation. We will study monotone rotations, where velocity $d\theta/d\tau$ has constant sign and nonzero integer average value $b$; see Fig. 4(b) and (c). In order to describe resonance rotations of the PPVL we will use the averaging method [Bogolyubov, Mitropolsky 1961] which requires rewriting (4) in the standard form as a system of first order equations with small right hand sides. For that reason we assume that $\varepsilon$, $\beta$ and $\omega$ are small positive parameters, which makes the system quasi-linear.

We introduce a vector of slow variables **x** and the fast time $s = |b|\tau$, where $x_1 = \theta - b\tau$ is the *phase mismatch*, $x_2 = d\theta/ds$ is the velocity, $x_3 = 1 + \varepsilon\cos(s/|b|)$ is the excitation. From here the dot denotes derivative with respect to new time $s$. Thus, Eq. (4) takes the standard form

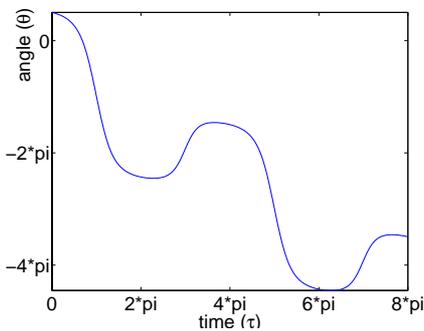

(a)  $\varepsilon = 0.51$, $\omega = 0.54$, $\beta = 0.05$.

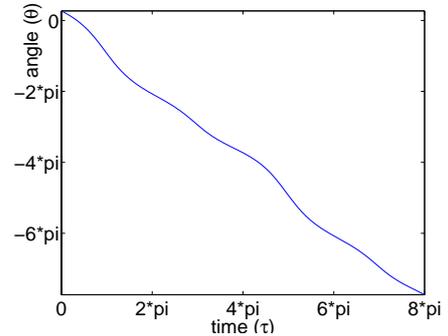

(b)  $\varepsilon = 0.28$, $\omega = 0.5$, $\beta = 0.05$.

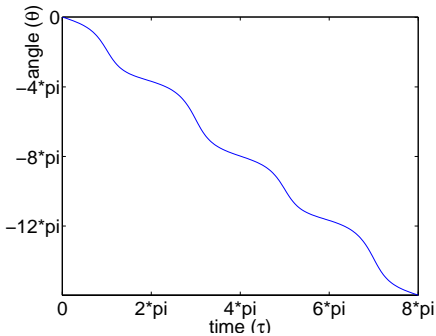

(c)  $\varepsilon = 0.43$, $\omega = 0.5$, $\beta = 0.05$.

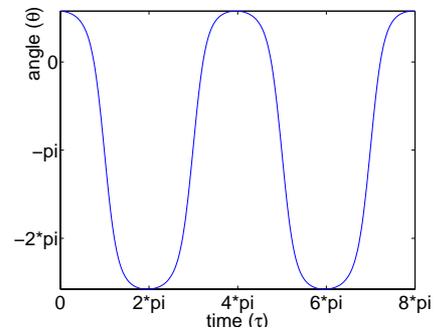

(d)  $\varepsilon = 0.59$, $\omega = 0.6$, $\beta = 0.05$.

**Fig. 4. (a) Regular rotation-oscillation with the mean angular velocity equal to one half of the excitation frequency, $b = -1/2$. (b) Regular rotation with $b = -1$. (c) Regular rotation with $b = -2$. (d) Regular rotation-oscillation with $b = 0$.**

$$\dot{x}_1 = x_2 - \mathrm{sign}(b),$$

$$\dot{x}_3 = \left(\varepsilon \frac{2\sin(s/|b|)}{x_3} - \beta\omega\right)\frac{x_2}{|b|}$$

(15)
$$- \frac{\omega^2}{b^2}\frac{\sin(x_1 + s\,\mathrm{sign}(b))}{x_3},$$

$$\dot{x}_3 = -\frac{\varepsilon}{|b|}\sin(s/|b|),$$

where it is assumed that $x_2 - \mathrm{sign}(b)$ is of order $\varepsilon$, $\mathrm{sign}(b) = 1$ if $b > 0$ and $\mathrm{sign}(b) = -1$ if $b < 0$. With the general averaging method we can find the first, second and the next order approximations of Eq. (15).

Resonance rotation domains of the PPVL for various $b$ are presented in Fig. 5. We see that greater values of relative rotational velocities $b$ are possible for higher excitation amplitudes $\varepsilon$. Numerically obtained rotational regimes are depicted in Fig. 5 by color points in parameter space $(\varepsilon, \omega)$ with $\beta = 0.05$. Domains of these points are well bounded below by analytically obtained curves for corresponding $b$.

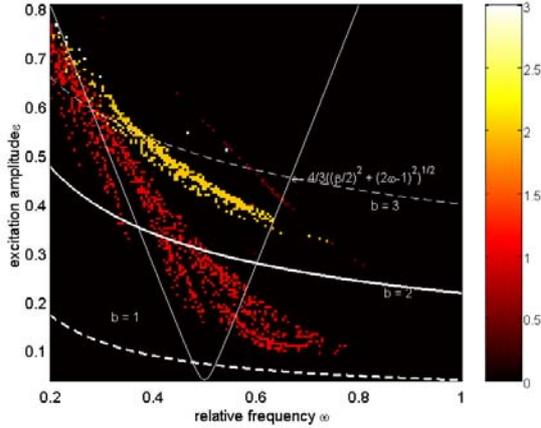

**Fig. 5.** Absolute values of relative rotational velocities are shown with different colors on the plane of parameters $\varepsilon$ and $\omega$ at damping $\beta = 0.05$. The correspondence between the colors and values is shown by the color bar on the right.

### 5.1. Rotations with relative velocity $|b| = 1$

It is the third order approximation of averaged equations where regular rotations with $|b| = 1$ under assumption that the coefficient $\varepsilon$ being of order $\omega^2$, and $\beta$ being of order $\omega^3$. Numerically obtained such rotations are shown in Fig. 4(b). In the third order approximation, averaged equations take the following form

(16)
$$\dot{X}_1 = X_2 - b$$
$$\dot{X}_2 = -\frac{3\varepsilon\omega^2}{2}\sin(X_1) - \beta\omega X_2$$

where $X_1$ and $X_2$ are the averaged slow variables $x_1$ and $x_2$. Auxiliary variable $x_3 = 1 + \varepsilon\cos(s/b)$ has unit averaged $X_3 = 1$ and is excluded from the consideration. Excluding variable $X_2$ from the steady state conditions $dX_1/d\tau = 0$ and $dX_2/d\tau = 0$ in (16) we obtain the equation for the averaged phase mismatch $X_1$

(17)
$$\sin(X_1) = -b\frac{2\beta}{3\varepsilon\omega}.$$

Thus, it is clearly seen from (17) that Eq. (16) has a steady solution only if

(18)
$$\omega \geq \frac{2\beta}{3\varepsilon}.$$

Inequality (18) gives the equation for the boundary of the domain in parameter space, where rotations with $|b| = 1$ can exist. This boundary is depicted with a bold dashed line in Fig. 5 on the parameter plane $(\varepsilon, \omega)$ for $\beta = 0.05$.

Stability of the solutions obtained from (17) was studied in [Seyranian, Belyakov 2008, Belyakov et al. 2008]. There was found the following condition for asymptotic stability

(19)
$$\cos X_1 > 0.$$

Hence, if inequality (18) is strict, then steady solutions

(20)
$$X_{1(1)} = -b\arcsin\left(\frac{2\beta}{3\omega\varepsilon}\right) + 2\pi k$$

are asymptotically stable, while solutions

(21)
$$X_{1(2)} = \pi + b\arcsin\left(\frac{2\beta}{3\omega\varepsilon}\right) + 2\pi k$$

are unstable, where $k = \ldots -1, 0, 1, 2\ldots$.

Thus, we conclude that if the parameters satisfy strict inequality (18) there are two stable regular rotations $\theta = b\tau + X_{1(1)}$ in opposite directions ($b = \pm 1$) and two unstable rotations $\theta = b\tau + X_{1(2)}$ in opposite directions.

### 5.2. Rotations with relative velocity $|b| = 2$

Rotations with higher averaged velocities $|b| = 2, \ldots$ correspond to higher excitation amplitudes $\varepsilon$. That is why here we will consider the coefficient $\varepsilon$ being of order $\omega$, and $\beta$ being of order $\varepsilon^3$. With this new ordering we obtain the sixth order approximation of the averaged equations for $|b| = 2$

(22)
$$\dot{X}_1 = X_2 - b/2$$
$$\dot{X}_2 = -\frac{9\varepsilon^2\omega^2}{16}\left[1 - \left(X_2 - \frac{b}{2}\right)^2 + \frac{\varepsilon^2}{27}\right]\sin(X_1) - \frac{\beta\omega}{2}X_2$$

which have steady state solutions determined by the following expressions

(23)
$$\sin(X_1) = -b\frac{4\beta}{9\varepsilon^2\omega}\left(\frac{1}{1 + \varepsilon^2/27}\right).$$

From Eq. (23) we get that the domain of rotations with $b = 2$ in the parameter space has the following boundary condition depicted in Fig. 5 with a bold solid line

(24) $\quad \omega \geq \dfrac{8\beta}{9\varepsilon^2}\left(\dfrac{1}{1+\varepsilon^2/27}\right).$

System (22) has similar structure to system (16). That is why stability condition for its steady state solutions appears to be the same as (19). Hence, if inequality (24) is strict, then solutions obtained from (23)

(25) $\quad X_{1(1)} = -\arcsin\left(\dfrac{4b\beta}{9\varepsilon^2\omega}\left(\dfrac{1}{1+\varepsilon^2/27}\right)\right) + 2\pi k$

are asymptotically stable, while solutions

(26) $\quad X_{1(2)} = \pi + \arcsin\left(\dfrac{4b\beta}{9\varepsilon^2\omega}\left(\dfrac{1}{1+\varepsilon^2/27}\right)\right) + 2\pi k$

are unstable, where $k = \ldots -1, 0, 1, 2, \ldots$.

Thus, as in the previous case, if the parameters satisfy strict inequality (24) there are two stable regular rotations $\theta = b\tau + X_{1(1)}$ in opposite directions ($b = \pm 2$) and two unstable rotations $\theta = b\tau + X_{1(2)}$ in opposite directions.

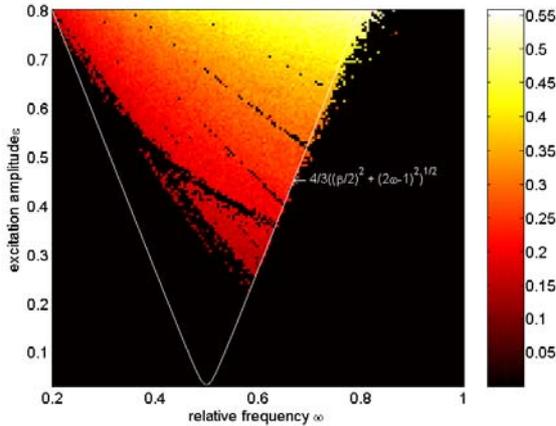

**Fig. 6. Maximal Lyapunov exponents are shown with different colors on the plane of parameters $\varepsilon$ and $\omega$ at damping $\beta = 0.05$. The correspondence between the colors and values is shown by the color bar on the right, where black color distinguishes the zero maximal Lyapunov exponent which corresponds to a regular regime while lighter colors correspond to positive exponents which characterize chaotic motions.**

## 6. Basins of attractions and transitions to chaos

In order to determine chaos domains we calculated maximal Lyapunov exponents presented in Fig. 6. We recall that positive Lyapunov exponents correspond to chaotic motions. Note that chaotic motion includes passing through the upper vertical position, i.e. irregular oscillations-rotations. This is usually called tumbling chaos.

We have observed two types of transition to chaos. The first type is when the system goes through the cascade of period doubling (PD) bifurcations occurring within the instability domain of the vertical position when the excitation amplitude $\varepsilon$ increases, for example at $\omega = 0.5$ in Fig. 7(a). The second type is when chaos immediately appears after subcritical Andronov-Hopf (AH) bifurcation when the system enters the instability domain of the lower vertical position of PPVL, for example at $\omega = 0.67$; see Fig. 7(b).

We can see the change of the system dynamics in its route to chaos along $\omega = 0.5$ in the bifurcation diagram shown in Fig. 7(a), where red points denote rotations with mean angular velocity equal to one excitation frequency ($|b| = 1$) and green points denote those equal to two excitation frequencies ($|b| = 2$). The domain with the most complex regular dynamics is surrounded by the red rectangle, where the system can have coexisting oscillations, rotations and rotations-oscillations.

Basins of attractions in Fig. 8 have been plotted using program *Dynamics* [Nusse, Yorke 1997]. These basins track the changes of the system dynamics in its route to chaos along $\omega = 0.67$. In Fig. 8(a) the oscillatory attractor (limit cycle) coexists with stationary attractor (lower vertical position of PPVL). In Fig. 8(b) we can see the first emergence of two rotational attractors with counterrotations. This picture is in a good agreement with condition (18) for existence of rotational solutions $|b| = 1$, see Fig. 5. Closer to the boundary of chaotic region in Fig. 8(c) only stationary and rotational attractors remain. Note that the basins of rotational attractors are small which means that at $\omega = 0.67$ the transition to chaos trough subcritical AH bifurcation is the most typical. In the middle of Fig. 8(d) the manifold of dark blue points reveals a typical strange attractor structure. The strange attractor inherits the basin of attraction from disappeared stationary attractor.

## 7. Conclusions

Asymptotic expressions for boundaries of the first instability domain of the lower vertical position have the form of half-cones in parameter space. Though, the second domain is empty. Approximate conditions for existence and stability of oscillatory and rotational motions are found analytically and compared with numerical study. Two types of transitions to chaos of the PPVL depending on problem parameters are distinguished. The first type is through the cascade of PD bifurcations. The second is through subcritical AH bifurcation. Different types of attractors with their basins where studied and depicted in Poincare section.

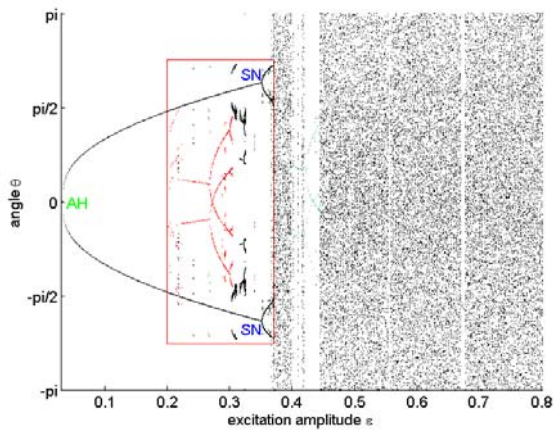 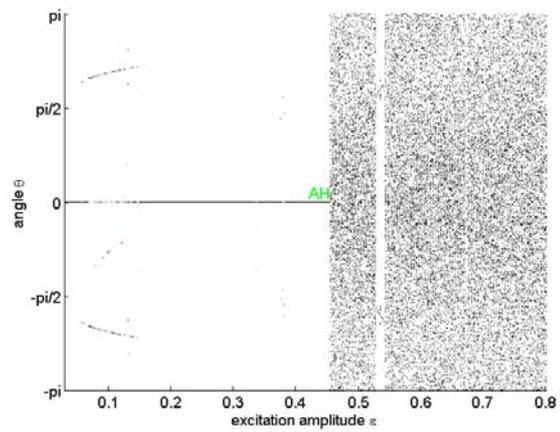

(a) $\omega = 0.5$.  (b) $\omega = 0.67$.

**Fig. 7.** The bifurcation diagram for different frequencies $\omega$ and the same damping $\beta = 0.05$ show two different types of transition to chaos. (a) After Andronov-Hopf bifurcation (AH) a limit cycle appears which experiences the saddle-node bifurcation (SN) and then the cascade of period-doubling bifurcations (PD). (b) After subcritical AH bifurcation of the vertical equilibrium the chaotic motion occurs immediately

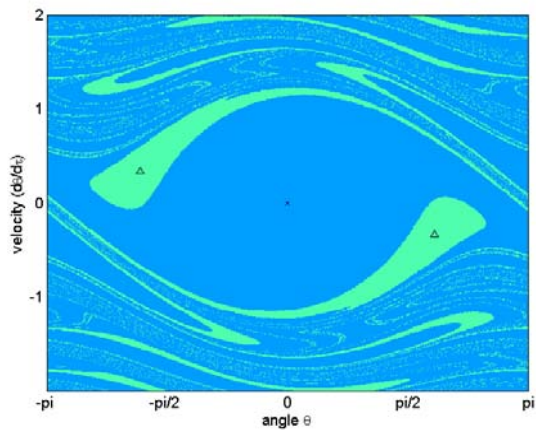 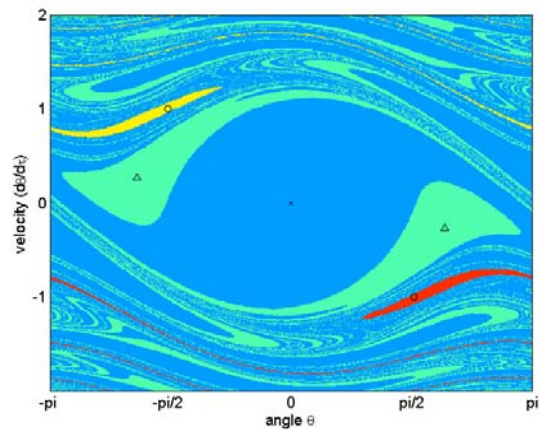

(a) $\varepsilon = 0.05$.  (b) $\varepsilon = 0.06$.

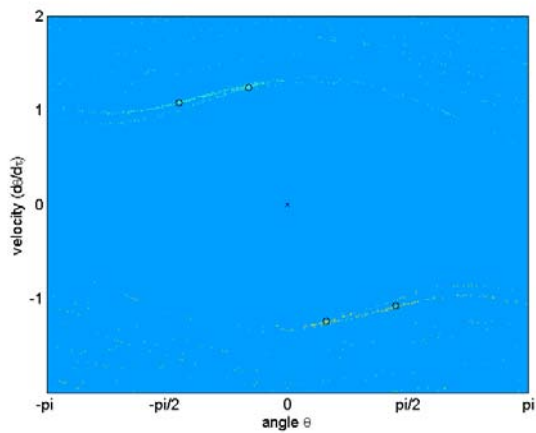 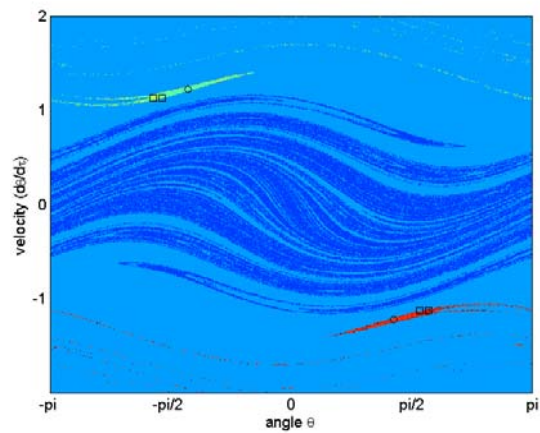

(c) $\varepsilon = 0.45$.  (d) $\varepsilon = 0.5$.

**Fig. 8.** Basins of attractions in Poincare section for different excitation amplitude $\varepsilon$ at the same frequency $\omega = 0.67$ and damping $\beta = 0.05$. △ marks period-two oscillational attractors, ○ marks period-one rotational attractors, □ marks period-two rotational attractors, × marks fixed points.

# НЕЛИНИЙНА ДИНАМИКА НА МАХАЛО С ПЕРИОДИЧНО ИЗМЕНЯЩА СЕ ДЪЛЖИНА

*Антон БЕЛЯКОВ, Александър СЕЙРАНЯН*

a_belyakov@inbox.ru,  seyran@imec.msu.ru

*Институт по механика, Московски държавен университет
Бул. Мичурински 1, Москва 119192, РУСИЯ*

*Изследвано е динамичното поведение на безтегловен прът с точкова маса, която се плъзга по пръта по периодичен закон. Това е махало с периодично изменяща се дължина, което се разглежда като прост модел на детска люлка. Изведени са асимптотични изрази за границите на неустойчивите области в близост до резонансните честоти. Аналитично са намерени области на трептеливи, въртеливи и трептеливо-въртеливи движения в параметричното пространство и са сравнени с числени изследвания. Числено са изследвани два типа преходи към хаос на махалото в зависимост от параметрите на задачата.*